\documentstyle[aps,prb]{revtex}
\begin{document}
\draft
\twocolumn[\hsize\textwidth\columnwidth\hsize\csname@twocolumnfalse\endcsname

\title{Large frequency range of negligible transmission in 1D photonic quantum well
structures}
\author{Jian Zi,\thanks{Electronic mail: jzi@fudan.edu.cn} Jun Wan and Chun Zhang}
\address{Surface Physics Laboratory (National Key Laboratory), Fudan University,\\
Shanghai 200433, People's Republic of China}
\date{\today }
\maketitle

\begin{abstract}
We show that it is possible to enlarge the range of low transmission in 1D
photonic crystals by using photonic quantum well structures. If a defect is
introduced in the photonic quantum well structures, defect modes with a very
high quality factor may appear. The transmission of the defect mode is due to
the coupling between the eigenmodes of the defect and those at the band
edges of the constituent photonic crystals.
\end{abstract}

] \newpage\narrowtext
\narrowtext

Since the work of Yablonovitch \cite{yab:87} and John \cite{joh:87} photonic
band-gap (PBG) materials have received considerable attention for
fundamental physics study as well as for potential applications in photonic
devices. \cite{sou:93,sou:95,joa:95} In PBG materials with any number of
dimensions, the dispersion relations (frequency versus wave-vector) possess
a number of branches. These branches form bands that might be separated by
frequency gaps owing to the periodic dielectric modulation, analogous to the
electronic band-gaps in semiconductors due to the periodic potentials.
Within a frequency gap electromagnetic (EM) waves cannot propagate. The idea
that 1D, 2D, and 3D periodic dielectric lattices can be designed to possess
PBGs has attracted wide interest, both theoretically and experimentally. 
\cite{sou:93,sou:95,joa:95} The absence of EM waves inside a PBG can lead to
some unusual features, which have many potential applications and novel
physics. By adjusting the geometric and dielectric parameters of a photonic
crystal, we can tailor the optic properties at will, the same way as in
semiconductor technology.

It is well known that many novel features for electronic properties appear
when two semiconductors form superlattice (SL) or quantum well (QW)
structures. In this letter, a transfer matrix method is used to study the
band structures of 1D periodic photonic structures and transmission of
photonic multilayers. We show that a large frequency range of very low
transmission exists in 1D photonic QW structures. By introducing a defect,
defect modes with very large quality factor may be induced.

Let us consider first the transmission and reflection of EM waves incident
on a single dielectric layer from the right and from the left. In general,
the solution of Maxwell's equations in layer $l$ can be written in the form

\begin{equation}
E(x|\omega )=E_{l1}e^{ik_lx}+E_{l2}e^{-ik_lx}.
\end{equation}
Here $k_l=\omega \sqrt{\epsilon _l}/c$ is the wave-vector in the $l$th layer
and $c$ and $\epsilon $ are the speed of light in vacuum and dielectric
constant, respectively. The coefficients $E_{l1}$ and $E_{l2}$ have to be
determined from the boundary conditions that both the electric field and its
first derivative are continuous across an interface. For convenience, the
field is written in the form of a vector as

\begin{equation}
E(x|\omega )=\left( 
\begin{array}{c}
E_{l1}e^{ik_lx} \\ 
E_{l2}e^{-ik_lx}
\end{array}
\right) .
\end{equation}
It can be shown that the field at $x_l$ in layer $l$ is related to the field
at $x_{l-1}$ in layer $l-1$ by a 2$\times $2 transfer matrix $T(x_{l-1},x_l)$

\begin{equation}
E(x_l|\omega )=T(x_{l-1},x_l)E(x_{l-1}|\omega ),
\end{equation}
where the transfer matrix is given by

\begin{equation}
T(x_{l-1},x_l)=P_l(\Delta x_l)Q_{l-1,l}P_{l-1}(\Delta x_{l-1}).
\end{equation}
Here $\Delta x_l=x_l-d_{l-1,l}$ and $\Delta x_{l-1}=d_{l-1,l}-x_{l-1}$ are
the distance from $x_l$ and $x_{l-1}$ to the interface between layers $l-1$
and $l$ located at $x=d_{l-1,l}$, respectively. It is easy to show that the
matrices $P$ and $Q$ are given by

\begin{equation}
P_l(\Delta x)=\left( 
\begin{array}{cc}
e^{ik_l\Delta x} & 0 \\ 
0 & e^{-ik_l\Delta x}
\end{array}
\right) ,
\end{equation}

\begin{equation}
Q_{l-1,l}=\left( 
\begin{array}{cc}
(1+\eta _{l-1,l})/2 & (1-\eta _{l-1,l})/2 \\ 
(1-\eta _{l-1,l})/2 & (1+\eta _{l-1,l})/2
\end{array}
\right) ,
\end{equation}
where $\eta _{l-1,l}=k_{l-1}/k_l$. The physical meaning of the two matrices
is that $P$ propagates the electric field a distance $\Delta x$ in a uniform
medium, whereas $Q$ makes the electric field from one side of an interface
to the other.

It is easy to show that for an $AB$ SL the band structures can be obtained
by applying the Bloch theorem

\begin{equation}
T(0,d)E(x|\omega )=E(x+d|\omega )=e^{ikd}E(x|\omega ),
\end{equation}
where $d$ and $k$ are the period and wave-vector of the SL, respectively.
The eigenfrequency is then obtained from

\begin{equation}
\cos (kd)=[T_{11}(0,d)+T_{22}(0,d)]/2,
\end{equation}
where $T_{ij}$ is the element of the 2$\times 2$ transfer matrix. The
transfer matrix for an $AB$ SL is given by

\begin{equation}
T(0,d)=Q_{B,A}P_B(d_B)Q_{A,B}P_A(d_A),
\end{equation}
where $d_A$ and $d_B$ are the width of $A$ and $B$ layer, respectively and $%
d=d_A+d_B$. After some algebra, it is easy to obtain the relation

\begin{eqnarray}
\cos (kd) &=&\cos (k_Ad_A)\cos (k_Bd_B)-  \nonumber \\
&&\frac{1}2\left( \eta _{A,B}+\frac 1{\eta _{A,B}}\right) \sin (k_Ad_A)\sin
(k_Bd_B).
\end{eqnarray}
From the above relation, photonic band structures can be obtained.

For a multilayer dielectric structure, the total transfer matrix is given by
\[
T^{(N)}=Q_{N,N+1}\prod_{l=N}^1P_l(d_l)Q_{l-1,l}, 
\]
where $N$ is the number of dielectric layers and $d_l$ is the width of the $%
l $th layer. When one encounters $N+1$ or $0$ in the subscript of the
matrices, the layer is treated as an air layer. From the total transfer
matrix, the transmissivity $t$ and reflectivity $r$ are given by

\begin{eqnarray}
t &=&\left| T_{11}^{(N)}-T_{12}^{(N)}T_{21}^{(N)}/T_{22}^{(N)}\right| ^2, \\
r &=&\left| T_{21}^{(N)}/T_{22}^{(N)}\right| ^2.
\end{eqnarray}
For lossless dielectrics, the condition $t+r=1$ must be satisfied.

In Fig. \ref{fig1} the band structures of photonic SLs, calculated by Eq.
(10), is given. The SLs consist of an air and a dielectric layer
alternatingly. PBGs appear in SLs due to the periodic modulation of the
dielectric constant. The two SLs given in the figure consist of the same
dielectric layers. The only difference is the filling factor, one (solid
lines) 0.5 and the other one 0.32. The second and third band of the $AB$ SL
is just inside the first and second PBG of the $CD$ SL, while the second
band of the $CD$ SL is just inside the second PBG of $AB$ SL. The
propagation of EM waves is forbidden in the PBGs. Therefore, we can use $AB$
and $CD$ multilayers to produce a QW structure, which may lead to a large
frequency range with very small transmission.

We display in Fig. \ref{fig2} the transmission of photonic QW structures $%
(AB)_m/(CD)_n/(AB)_m$, where $(AB)_m$ means that this multilayer consists of 
$m$ sublayers of $AB$. The transmission was calculated by Eq. (11). The $CD$
layers play a role of well or barrier depending on the frequency of the EM
wave. In the absence of $CD$ layers (Fig. \ref{fig2}a) the second and third
bands of $AB$ multilayers have large transmission. With the introduction of
a well or barrier layer $CD$, the transmission of these bands is suppressed
owing to the fact that these two bands are just inside the second and third
PBGs of a $CD$ multilayer. For a small number of $CD$ multilayers, there is
still some transmission for the second and third band of the $AB$
multilayer. With the increase in number of $CD$ layers, the transmission of
these bands is reduced rapidly. When the number of $CD$ layers is greater
than 10, the transmission is very small and can be neglected. In this case,
a large forbidden gap for transmission exists with a very large range in
reduced frequency from 0.16 to 0.72. With the photonic QW structures, by
properly choosing the geometric and dielectric parameters, a large forbidden
gap for transmission can be achieved, which may have potential applications.
For instance, this structure can be used as a nearly perfect mirror that can
reflect all visible light.

By introducing some defects in the photonic QW structures, defect modes can
be induced, as shown in Fig. \ref{fig3}. It can be seen from the figure that
the defect peak is very sharp with the quality factor as high as several
thousands. This defect QW structure can be used as a microcavity or as a
high quality filter. The very sharp transmissivities of the defect modes are
due to the coupling between the eigenmodes in defect layers and those at the
band edges of the $AB$ or $CD$ layers. Normally, the coupling between the
eigenmodes of defects and those away from the band edges is very small.
Therefore, in some cases, the transmission of a defect mode is very small if
the coupling is small. This behavior is somewhat different from that in the
usual dielectric multilayers.

In summary, 1D photonic QW structures have been suggested. By properly
choosing the geometric and dielectric parameters of the photonic
constituents, a large frequency range of negligible transmission of EM waves
is possible, which may have potential applications. By introducing some
defects in the photonic QW structures, defect modes with very large quality
factors may appear. The sharp defect transmission peaks are due to the
coupling between the eigenmodes of defect layer and those at the band edges
of the constituents.

This work is supported by the National Natural Science Foundation of China
under Contract No. 69625609.

\begin{figure}[htbp]
\caption{Photonic band structures of photonic SLs. The periodicity of the
SLs is $a$. Solid lines are for an $AB$ SL with sublayer width $d_A=d_B=0.5a$
and dielectric constant $\epsilon _A=1$, $\epsilon _B=13$. Dashed lines are
for an $CD$ SL with $d_A=0.68a$, $d_B=0.32a$ and $\epsilon _C=1$, $\epsilon
_D=13$.}
\label{fig1}
\end{figure}

\begin{figure}[htbp]
\caption{Transmission of photonic QW structures (AB)$_{10}$/(CD)$_n$/(AB)$%
_{10}$. (a) $n=0$, (b) $n=1$, (c) $n=2$, (d) $n=5$, (e) $n=10$.}
\label{fig2}
\end{figure}

\begin{figure}[tbp]
\caption{Transmission for photonic QW structures with defects. The QW
structure is $(AB)_{10}/(CD)_{10}/(AB)_{10}$ and the defect is inserted in
the middle of the $CD$ multilayers. (a) The defect is an $EF$ layer with $%
d_E=0.6a$, $d_F=0.4a$ and $\epsilon _E=1$, $\epsilon _F=13$. (b) The defect
is a single layer with $d=0.5a$ and $\epsilon =1$.}
\label{fig3}
\end{figure}


\begin{references}
\bibitem{yab:87}  E. Yablonovitch, Phys. Rev. Lett. 58, 2059 (1987).

\bibitem{joh:87}  S. John, Phys. Rev. Lett. 58, 2486 (1987).

\bibitem{sou:93}  {\it Photonic Band Gaps and Localization, NATO ARW}, ed.
by C. M. Soukoulis (Plenum Press, New York,1993).

\bibitem{sou:95}  {\it Photonic Band Gap Materials, NATO ASI}, ed. by C. M.
Soukoulis (Kluwer Academic Publishers, Dordrecht, 1996).

\bibitem{joa:95}  J. D. Joannopoulos, R. D. Meade, and J. N. Winn, {\it %
Photonic Crystals} (Princeton Univ. Press, Princeton, 1995).
\end{references}
\end{document}